\documentclass[12pt,preprint]{aastex}

\def\kms{\relax \ifmmode {\,\rm km\,s}^{-1}\else \,km\,s$^{-1}$\fi}

\def\Mso{{M$_{\rm \odot}$}}
\def\Mj{{M$_{\rm J}$}}
\def\Rso{{R$_{\rm \odot}$~}}
\def\Lso{{L$_{\rm \odot}$}}

\def\cm3{${\rm cm}^{-3}~$}

\slugcomment{}

\shorttitle{CAN PLANETS SURVIVE STELLAR EVOLUTION?}
\shortauthors{Villaver \& Livio}
  
\begin{document}
  
\title{CAN PLANETS SURVIVE STELLAR EVOLUTION?}
\author{Eva Villaver\altaffilmark{1} and Mario Livio}
\affil{Space Telescope Science Institute, 3700 San Martin Drive,
Baltimore, MD 21218, USA; villaver@stsci.edu, mlivio@stsci.edu}

\altaffiltext{1}{Affiliated with the Hubble Space Telescope Division
of the European Space Agency} 

\begin{abstract}

We study the survival of gas planets around stars with masses in the range
1--5 \Mso, as these stars evolve off the Main Sequence. We show that planets
with masses smaller than one Jupiter mass do not survive the Planetary Nebula
phase if located initially at orbital distances smaller than (3--5)
\,AU. Planets 
more massive than two Jupiter masses around low mass (1 \Mso~on the Main
Sequence) stars survive the Planetary Nebula stage down to orbital distances
of $\sim$3 \,AU. As the star evolves through the Planetary Nebula phase, an
evaporation outflow will be established at the planet's surface. Evaporating
planets may be detected using spectroscopic observations. Planets around
white dwarfs with masses M$_{WD}\gtrsim$ 0.7 \Mso~are generally expected to
be found at orbital radii r $\gtrsim$ 15\,AU. If planets are found at smaller
orbital radii around massive white dwarfs, they had to form as the result of
the merger of two white dwarfs.  

\end{abstract}


\section{INTRODUCTION}
In recent years, the quest for Jupiter-like giant planets has been
extended to a completely different kind of hosts--white dwarfs 
(e.g.~\citealt{Detal:02,Farihi:05,Fetal:05}). The underlying assumption has
been that 
planets can survive the parent star's 
evolution. Once this is accepted, the fact that white dwarfs are $10^3$ to
$10^4$ times fainter than their main sequence progenitors opens up the
possibility to observe planets through direct imaging in the infrared
\citep{Bch:02} where the planet emission peaks \citep{Betal:97,Ig:01}.

In general, it has been assumed that planets survive to the white dwarf stage
if they manage to stay in a large enough orbit to avoid engulfment by the star
when the latter increases its radius as it ascends the Red Giant Branch (RGB)
and Asymptotic Giant  Branch (AGB). Therefore, studies of the planet's fate as
the star evolves off the main sequence have generally been restricted to
determining the planet's orbit
(e.g. \citealt{Ls:84,Sok:94,Sbk:93,Retal:96,Rd:01,Bch:02}) or the orbit's 
stability (e.g. \citealt{Dl:98}). However, before the star reaches the white
dwarf stage, 
and immediately after the AGB evolution, the star evolves into the planetary
nebula (PN) phase, where the stellar temperature can reach 300\,000 {\rm K}
with 
luminosities of $10^3$ \Lso, while emitting powerful winds. 

In this paper we explore the effects of an evolving AGB and post-AGB star
on an orbiting planet. In \S2 we examine the planet survival during the AGB
phase and the orbital changes due to the AGB mass-loss. In \S3 we
estimate the thermal conditions at the planet's surface as the star increases
its temperature during the PN phase. In \S4 we estimate the range of
parameters under which an outflow from the planet will ensue, and we
estimate the planet's 
evaporation rate due to thermal heating. In \S5 we qualitatively discuss the
planet's reaction to mass-loss and other processes that might influence the
planet's evolution, and in \S6 we discuss the planet's overall survival. Our
conclusions follow.

\section{PLANET SURVIVAL DURING THE AGB PHASE} 

Single white dwarf progenitors are expected to have main sequence masses in the
range from 1 to about 8 \Mso~mass (the upper mass limit is not well
established). As these stars leave the main sequence they evolve into 
the RGB, Horizontal Branch, AGB, and PN phases in the Hertzsprung-Russell (HR)
diagram, before  descending the  white dwarf cooling track. The major
structural changes 
in the post-main sequence evolution of low- and  intermediate-mass stars occur
during the RGB and AGB phases. During the RGB and AGB the stellar effective
temperature 
is always lower than its main sequence value and therefore it has no influence
on the planet's survival. However, it is during the late AGB evolution, the
so-called  thermal-pulsing AGB phase, that a planet's orbit will be most
influenced, since during this phase the star loses most of its initial mass and
reaches its maximum radius.

The planet will spiral-in and  evaporate totally (or in rare cases will
accrete mass and become a 
close, low-mass companion to the star) if the planet's orbital distance is
within the 
reach of the star's radius during the AGB phase \citep{Ls:84}. An estimate of
the maximum planet mass that can be evaporated inside an AGB envelope can be
obtained by  equating the location of the evaporation region (where the local
sound speed in the stellar envelope matches the escape velocity from the
planet's surface) to the energy required to expel the envelope
\citep{Sok:96,Sok:98,Nt:98}.  The value of this maximum mass is very uncertain
because it depends on several factors, such as the efficiency of envelope
ejection \citep{Pzy:98}, which are
largely unknown.  Using the simplified formalism of \cite{Nt:98} we find
that  planets with masses less than  0.014 \Mso~or 15~\Mj~(where \Mj~is the
Jupiter mass) will evaporate inside the envelope of an AGB star with  main
sequence mass of 1~\Mso. This mass limit is much higher, $\sim$ 120 \Mj (0.11
\Mso) well into the stellar regime, if the planet (or brown dwarf) is engulfed
inside the AGB envelope of a 5 \Mso~star.  

We can therefore safely assume that planets less massive than 15 \Mj~that are
engulfed, will be dissipated, 
because this limit corresponds to the temperature (and therefore
planet's 
escape velocity) reached inside the AGB envelope for the lowest mass
stars (in the range examined here). Note that this planet mass limit 
is well below the mass of the brown dwarf (52 \Mj) recently discovered by
\cite{Max:06} orbiting an under-massive white dwarf. On the other hand, the
case of total planet evaporation (for a planet in a close orbit around a
solar-like star) has been proposed to explain the formation of 
single under-massive white dwarfs \citep{Nt:98}.

The structural changes that an AGB star undergoes in response to the
dissipation of a planet in its interior are complex and have been extensively
explored by \cite{Sl:99a,Sl:99b} and by
\cite{Setal:02,Setal:04}. The details of the destruction
of such a planet within the stellar envelope of an AGB star are beyond the
scope of the present work. 

\subsection{The Evolution of the Planet's Orbit}

In the following we concentrate on Jupiter-like planets in orbits that {\it
  avoid} 
engulfment when the radius of the AGB star expands during the thermal pulses,
and we explore the conditions under which the planet will remain outside the
stellar surface.  

For orbital distances larger than the stellar radius, $R_*$, there is still a
range of orbits 
for which drag effects (gravitational and tidal) are important in decreasing
the planet's orbit, ultimately causing the planet to spiral into the AGB
envelope. While the gravitational drag, caused by the increase in the planet
mass through the accretion of material from the stellar wind, is negligible for
Jupiter-like planets \citep{Dl:98}, the decrease in the planet's
angular momentum caused by the tidal drag cannot be neglected for orbits close
to the stellar radius during the AGB phase. 

If the planet's orbital distance
is larger than the stellar radius reached when the star expands during the AGB
phase, the variation of the orbit can be approximated by
(e.g.~\citealt{Zah:77}):  

\begin{equation}
\frac{1}{r} \frac{dr}{dt}=-\frac{1}{M_*} \frac{dM_*}{dt}+(\frac{1}{r}\frac{dr}{dt})_{tidal}
\end{equation}

\noindent
where $r$ is the orbital distance and $M_*$ is the stellar mass on the main
sequence. The first 
term on the right hand side of Eq.~1 represents the increase in the orbital
radius due to mass-loss from the star and the second 
term represents the decrease due to tidal interaction, which is 
proportional to $[R_*(t)/r]^8$ \citep{Zah:77} and becomes negligible at large
orbital distances. At small orbital 
distances, however, the decrease in the planet's angular momentum caused by the
tidal interaction can lead to a decrease in the orbit by $\sim$ 100 to 300 \Rso
\citep{Ls:83,Retal:96,Rd:01}, depending on the planet's mass and the ratio of
the rotational to orbital angular velocity. Note however that the stellar
radius during the AGB increases gradually. Variations in the surface
luminosity and radius of an AGB star arise as a 
consequence of the thermal and structural readjustments produced during the
after-pulse phase. In particular, $R_*$ is not maintained after a
thermal-pulse. Rather, 
the star eventually contracts as the temperature in the He-burning shell
decreases and a new thermal pulse begins. 

The maximum stellar radius,
$R_*^{max}$, is only
reached briefly at the end of the AGB evolution as the star gets to the  
AGB tip luminosity. Consequently, it is safe to neglect 
the tidal term in Eq.~1 for orbital radii larger than $R_*^{max}$. The AGB
and  post-AGB evolution of the stellar mass-loss 
rate, effective temperature, and luminosity that we use have been  obtained
from 
the stellar evolution models of \cite{Vw:93, Vw:94}. From  these, we have
calculated the maximum radii , $R_*^{max}$, reached by stars of different
masses during the AGB phase. These values of  $R_*^{max}$ are listed in
Table~1. 
Neglecting tidal
interaction for $r \ge R_*^{max}$ and integrating Eq.~1, the
orbital radius is then given by
\begin{equation}
r(t)=r_o \frac{M_*}{M_*(t)}
\end{equation}
\noindent
where $r_o$ is the initial orbital radius and $M_*(t)$ is time-dependent
stellar mass caused by mass-loss.  

In Fig.~1 we show the evolution of the orbital radius during the late AGB phase
by using $r_o~=~R_*^{max}$ for each initial mass considered. In the
calculation of  $M_*(t)$ we have considered the fact that the AGB mass-loss
accounts for 
most of the mass lost by low-and intermediate-mass stars and that mass-loss
previous to the AGB is negligible. This is a consequence of the fact that
mass-loss rates on the RGB are proportional 
to $R_*L_*/M_*$ \citep{Rei:75}, resulting in a significant amount of envelope
mass lost before the AGB phase only for stars with initial masses
$M_*~<$~1~\Mso.   
 
The higher the initial mass, the higher the amount of mass lost during the AGB
phase and hence the larger the planet's orbital expansion. Planets
reach a final orbital distance at the end of the AGB phase,
determined by multiplying the initial orbit by  $M_*/M_{WD}$ where $M_{WD}$
is the white dwarf mass, given that
mass-loss is negligible once the star leaves the AGB. The orbital expansion
factors for each stellar mass are given in column (4) of Table 1. 

Another effect to consider is whether the planet
becomes  unbound due to the change in mass of the central star.  Unbinding can
be expected if the stellar mass-loss timescale, $\tau_{mass-loss}$, satisfies
$\tau_{mass-loss} < \tau_{dyn}$, where $\tau_{mass-loss}$ is given by
\begin{equation}
\tau_{mass-loss} \sim \frac{M_*}{\dot{M}_*}
\end{equation} 
and the dynamical timescale,$\tau_{dyn}$, by
\begin{equation}
\tau_{dyn} \sim [\frac{r^3}{G(M_*+M_{p})}]^{1/2},
\end{equation}
with $\dot{M}_*$ being the stellar mass-loss rate, $M_p$ the planet's mass,
and G the gravitational constant.  
Given that $\tau_{dyn}\sim$50\,yr while the shortest mass-loss timescale is
$\tau_{mass-loss} \sim 10^5$\,yr, it is very unlikely that a planet will
become unbound due to the decrease in the stellar mass during the AGB phase. 
 
A gas planet with $M_p~<~$15 \Mj~orbiting an AGB star (of any mass) at an
initial orbital  distance $r_o~<~R_*^{max}$ will most likely evaporate inside
the 
stellar envelope. The deuterium  burning minimum mass limit (12 \Mj) is
traditionally used  as the mass boundary between planets and brown dwarfs
(e.g. \citealt{Setal:96}). Therefore we can safely state that gas planets
do not 
survive the AGB 
evolution if their  initial orbits lie within the reach of the stellar radius
during the AGB evolution, whereas brown dwarfs can survive.  For
$r_o~\ge$~R$_*^{max}$ the orbit will generally expand due to  the heavy stellar
mass-loss rates experienced during the AGB evolution. Larger differences
between the initial and final mass of the star are experienced for the more
massive progenitors, causing the orbits of planets orbiting the more massive
stars (note that we are always refering to stars in the 1--5 \Mso~mass
range for which complete stellar evolution models exist) to be modified by
the larger factors (up to 5.5 times larger than the initial orbit). 

\section{THE CHANGE IN THE PLANET'S ENVIRONMENT DURING THE PN PHASE}  

During the post-main sequence evolution of the star the stellar effective
temperature 
always remains lower than its main sequence value. However, once the star
leaves the AGB phase, the high   mass-loss rate ceases and the remnant core
moves in the HR diagram  at constant luminosity toward higher effective
temperatures into the PN stage, before the star reaches the white dwarf cooling
track. The planet's orbit is not expected to change further at this stage.
However, the main processes responsible for shaping PNe (high velocity winds)
and powering PNe line emission (high stellar effective temperatures) need to be
considered to establish the survival of a planet during this phase.

The luminosity, mass, and timescale of evolution of the star during the PN
phase depend mostly on the stellar core mass
\citep{Pac:71,Ir:83,Vw:94, I:95}. The stellar luminosity during this phase is
within the range 3.5 to 23 $\times 10^3$ \Lso~(for the lowest 0.56 and highest
0.9 \Mso~mass remnant, respectively) and the stellar temperature can reach
100\,000--380\,000\,K (for the same core masses, respectively). The hydrogen
ionizing photon flux, which is of the order of 10$^{48}$~s$^{-1}$
\citep{Vmg:02}, is 
responsible for the PNe  
ionized line emission. PNe central stars also emit very high velocity winds
(with speeds of a few thousands of \kms) which are driven by  the transfer of
photon momentum to the gas, through absorption by strong resonance lines (e.g.
\citealt{Petal:88}). The PN is largely shaped by the  interaction of this high
velocity wind with the slowly ejected material during the AGB phase
\citep{Kpf:78,Kw:85,Balick, Mf:95,Vgm:02,Vmg:02}.  The survival of a gas
planet as the star evolves into the  PN phase strongly depends on the planet's
surface temperature,
which ultimately determines whether or not high evaporation rates are set at
the planet's surface.

\subsection{The Planet's Equilibrium Temperature}

The radiative equilibrium temperature of a planet can  be estimated by
balancing the flux received by the star to the blackbody flux re-radiated
by the planet, 

\begin{equation}
\sigma T_p^4=\frac{L_*}{16 \pi r^2} (1-A)
\label{Teq}
\end{equation}

\noindent
where $\sigma$ is Steffan-Boltzmann's constant, $T_p$ is the planet's
equilibrium temperature, $r$ is the orbital radius, $L_*$ is the stellar
luminosity 
and $A$ is the Bond albedo. We have assumed  an approximate value of
$A=$0.5. As shown in \cite{Marley} Bond albedos show little sensitivity to
the planet's 
gravity or effective temperature for a given stellar spectrum; the largest
variations arise from the stellar spectral type and the presence of
clouds. 

Fig.~\ref{teq_fig} shows the planet's equilibrium temperature versus orbital
distance as given by Eq.~\ref{Teq} for the two extreme cases of 
primaries: the lowest and the highest stellar masses
considered. Internal energy sources have not been considered in
Eq.~\ref{Teq} since they are expected to be negligible.

A planet orbiting a PN central star will have an age in the range  0.1 to 10
\,Gyr as the star enters the PN phase, assuming that the star and the planet
are coeval. \cite{Bur:01} and \cite{Hub:02} show the
predicted range of effective temperatures for {\it isolated} giant
planets and brown dwarfs of various masses to be between 100\,K and 400\,K for
the age range of interest here. The cooling theory applies equally to the
structure and evolution of brown dwarfs and giant planets, given their
convective nature. From Fig.~\ref{teq_fig}  we see that the planet's
equilibrium temperature in the presence of the PN core is much higher than
the expected value in isolation, at 
orbital distances $r~\le$ 20 \,AU and $r~\le$ 50 \,AU for the 0.56 and 0.9
\Mso~stars respectively.

\subsection{The Planet's Exospheric Temperature} 

Given the high effective temperature of the star, most of the stellar flux is
emitted at short wavelengths, for which the material in the planet's
atmosphere has a high optical depth. The incident flux will therefore be
absorbed 
by a thin surface  layer, the depth of which will be determined by the
transparency of 
the material to the incident photons. Inward to this surface layer, the heat
transfer to the interior is set by a combination of conduction and convection.
The temperature on the planet's atmosphere has to be determined through the
energy equation considering  radiative, conductive and convective
terms.  Convective effects are expected to be important in the atmosphere of a
tidally synchronized planet, since a strong temperature gradient will be set
between the day and night side of the planet. However, only very short period
planets 
are expected to be tidally synchronized \citep{Gui:96} and planets orbiting
PN central stars are not expected to be found at such short orbital periods
(see \S2). Therefore, it is reasonable to assume that, for the case under
consideration, convection can be neglected as first approximation.
Moreover, it has been shown that in the upper atmosphere of a planet (the
exosphere) the density becomes very low and the temperature is determined
mainly by 
the absorption of extreme ultraviolet and X-ray (XUV) radiation 
\citep{Bl:04}. Therefore by neglecting  convective effects the
energy equation takes the simplified form: 

\begin{equation}
{\nabla}\cdot (K\,{\nabla}T-{q}_{\rm xuv})=0 
\label{Energy}
\end{equation}

\noindent
where ${q}_{\rm xuv}$ is the effective radiative heat flux of the
star at wavelengths shorter than 3000\AA~for which the opacity is very high,
$T$ is the temperature, and $K$ is the thermal conductivity. The exospheric
temperature, $T_{ex}$, can be obtained  by 
integrating Eq.~\ref{Energy}. This gives (see e.g. \citealt{Bau:71, Bau:73}), 
\begin{equation}
T_{ex}^{s+1}-T_b^{s+1}\simeq\frac{ \alpha k_b \sigma_c \epsilon I_{\rm \infty} 
  }{K_o m_H g \sigma_a} T_{ex}
\label{tex}
\end{equation}

\noindent
where $s$ is the temperature dependence of the thermal conductivity,
$s$~=~0.7 
for atomic hydrogen \citep{Ds:62}, $T_b$ is the temperature at the base of
the thermosphere, $\alpha$ is a factor  that
takes into account the rotation of the planet and varies between 0.5 and 0.25
for rapid or slow rotators, $k_b$ is Boltzman's constant, $\sigma_c$ is the
gas-kinetic collision cross-section,  $K_o$ is the thermal conductivity
proportionality 
factor,  $m_H$ the mass of the hydrogen atom, and $g$ is
the gravitational  acceleration.
The intensity of the 
stellar radiation reaching the planet's
atmosphere $I_{\rm \infty}$ is related to ${q}_{\rm xuv}$ by,
\begin{equation}
{q}_{\rm xuv}= n_j \sigma_a\epsilon I_{\rm \infty} e^{-\tau} 
\end{equation}
where  $\epsilon$ is the heating
efficiency (the fraction of the stellar 
flux absorbed that is transformed into thermal energy in the atmosphere), and
$\tau$ is the optical depth, with $\tau= \int n_j \sigma_a dz$ where $n_j$ is the
number density 
of the absorbing constituent (atomic hydrogen), $\sigma_a$ its absorption
cross section to XUV radiation and $z$ is the atmospheric altitude. 

Since $T_b$ is roughly
equal to the planet's equilibrium temperature,  then $T_b \ll T_{ex}$ and the
second term on the left hand side of Eq.~\ref{tex} can be neglected.
Note that in order to estimate $T_{\rm ex}$ in Eq.~\ref{tex} cooling in the
infrared has not been considered. This is justified by 
the fact that most of the molecular hydrogen (which is responsible for the
cooling in the infrared) is expected to be dissociated in
the upper atmosphere.  

In Fig.~\ref{exo_temp} we show the exospheric temperature $T_{\rm ex}$
reached by a Jupiter-like planet of 1 \Mj~mass due to the XUV generated by a
0.56 \Mso~(1 \Mso~main 
sequence mass) central star versus the orbital distance. The different curves
are the result of changing the XUV flux intensity due to the evolution of the
star in the HR diagram. The stellar luminosity emitted below 3\,000\AA~has
been estimated using the GMFGEN code of \cite{Hill:98,Hill:99}. Note
  that for
 a PN central star with an effective temperature of  115\,000 \,K
(and log g =   
5.7), 97 \% of the total stellar luminosity is emitted below 3\,000~\AA. For
cooler central stars (e.g. 35\,000 \,K), 93\% of the stellar luminosity is 
below 3\,000 \AA. The solid 
curve represents $T_{\rm ex}$ obtained by using the stellar flux  
early in the evolution of the star into the PN phase when the stellar
luminosity is $L_*$~=~3\,600 \Lso~and $T_{\rm eff}$~=36\,000\,K; the
dashed curve corresponds to an intermediate stage  ({\rm $L_*$}~=~1\,400
\Lso~and $T_{\rm eff}$~=115\,000\,K); and the dotted curve corresponds to a
point 10$^5$\,yr later in the evolution, when the luminosity has dropped to
100 \Lso~at a temperature of $T_{\rm eff}$~=96\,000\,K.  
In the following, we explore the conditions under which an evaporation flow is
established at the planet's surface.

\section{PLANET EVAPORATION DUE TO THERMAL HEATING DURING 
  THE PN PHASE}

The standard formulation for the gas-kinetic or thermal evaporation from
celestial bodies was first established by \cite{Jea:25}.  Jeans's formula gives
a lower limit to the rate of escape (by thermal evaporation) of gas from a
planetary atmosphere: 

\begin{equation}
\Phi_J = \frac{v_o}{2\sqrt\pi} n_c (1+E) e^{-E}
\label{jeans}
\end{equation}

\noindent
where $v_o = \sqrt{2 k_b T/m_H}$ is the velocity of the escaping particles, 
$n_c$ is the number density of the escaping constituent (we assume 
hydrogen from now onwards) at a given critical level, $T$ is the temperature, 
and $E$ is the
escape parameter $E(R)~=~G M m_H /R k_b T$ where $M$ is the planet's mass
contained within a spherical surface of radius $R$, and $k_b$ and $G$ are
Boltzman's and the gravitational constants respectively.
{\it Jeans escape} then usually refers to the
density of the critical level $n_c$ at the exobase, that delineates 
the lower limit of the exosphere (e. g.~\citealt{Ch:87}). The temperature of this
layer, the exospheric temperature T$_{\rm ex}$, is much  higher than the
planet's effective  temperature (given by Eq.~\ref{Teq}),  and higher than the
gas photoionization temperature (10\,000 \,K) and is the
characteristic planetary temperature for thermal escape
\citep{Mou:01,Letal:03,Betal:04, Letal:04}.   A common misconception when
considering the  evolution of a planet 
inside a PN, is that the evaporation rates are determined by the Jeans
escape, with 
a temperature of 10\,000 \,K fixed by photoionization. Using the
ionization temperature the sound speed of the
ionized gas in the planet's atmosphere is  smaller than the typical escape
velocity 
for the material and therefore no evaporation is expected. 

Equation~\ref{jeans} seems to imply that high evaporation rates may be
achieved for high values of  $n_c$. However, {\it Jeans escape} is only valid when
the escape rate is relatively small and therefore the static structure of the
atmosphere is not perturbed, that is, when the escape parameter $E\ge$~20-30. 
Note that $E$ represents the ratio of the gravitational potential energy to the
particle kinetic energy or, in other words, the ratio of the square of the
escape to the 
thermal velocity in the upper atmosphere. For large escape fluxes, $E\le$20, the
structure of the upper atmosphere is modified, the gas that has escaped has to
be replaced by gas from the lower levels thereby altering the atmospheric
structure. In this case, the escape rate given by Eq.~\ref{jeans},  and the
thermal balance 
are no longer valid \citep{Ch:87, Bl:04}. For $E \le E_c$, where
$E_c\approx 1.5-3$, the atmosphere reaches blow-off conditions
(e.g.\citealt{Opi:63}). The exospheric base is
lifted outwards, the planet radius increases and the atmosphere is lost if the
heat source continues to operate. Under blow-off conditions the planet's
atmosphere will expand radially outward at relatively modest, subsonic
velocities, close to the surface, and will gradually accelerate to  supersonic
velocities as it moves further away.
Fig.~\ref{exo_temp} shows that blow-off conditions (represented by the region
above the solid or the dotted horizontal line depending on the $E_c$ value
assumed) are easily reached  at the planetary atmosphere up to distances of
125 \,AU.  

As the stellar mass is increased, higher luminosities and effective temperatures
than the ones used in Fig.~\ref{exo_temp} are reached during the PN phase. More
intense  XUV radiation fields will reach a planet's atmosphere orbiting a
higher mass progenitor. The results  shown in Fig.~\ref{exo_temp} therefore
represent a  lower limit (in stellar mass) to the development of blow-off
conditions.  Although we have assumed a hydrogen composition to compute 
$T_{\rm ex}$,  if other, heavier constituents exist they will be carried away
through drag 
forces (by the light gas moving at a sonic speed) once the atmosphere reaches the
blow-off  condition.

The escape parameter, as well as the exospheric temperature, depend on the
planet's surface gravity. The higher the planet's mass, the harder it is to
reach the blow-off conditions because  the particles need a much higher thermal
energy to reach escape velocities at the planet's surface. The gray area in
Fig.~\ref{exo_mass} shows the orbital distance region (to 125 \,AU)  at which a
planet's or a brown dwarf's atmosphere will reach blow-off conditions as a
function of the 
planet (or brown dwarf) mass. In the calculation we have used mass-radius
relations given by \cite{Betal:03} for non-irradiated extrasolar planets at  10
\,Gyr, and assumed that the exospheric temperature is given by Eq.~\ref{tex}.
The left and right panels have been computed for a 0.56 \Mso~(1
\Mso~main sequence) and 0.9 \Mso~(5\Mso~main sequence) star respectively,
early in the PN phase. We have chosen 
the 
same effective temperature of the star, 36\,000\,K, which corresponds to
$L_*$~=~3\,600 \Lso~and $L_*$~=~23\,000 \Lso~respectively.  Higher mass stars
produce more intense XUV radiation fields and blow off 
conditions are set at larger orbital distances for a given planet mass. The
evolution of the star towards higher effective temperatures has similar 
effects. 

\subsection{The Planet's Evaporation Rates}

Given the high exospheric temperatures reached at the planet's upper atmosphere
it is expected that an outflow will develop as a consequence of the
absorption of the XUV radiation. Since high temperatures can cause the
outer layers of the planet to escape rapidly, in order to examine the planet
survival, we have to estimate the evaporation rate. Note that at the orbital
distances we are considering, the planet will be well within the typical inner
radius of the nebular shell (0.01 to 0.1 \,pc) \citep{Vmg:02}, so we do not
expect a decrease in the photon flux arriving at the planet's surface due to
absorption by the nebula. 

The problem of a general outflow from a stellar (or planetary) body can be 
described with the same set of equations used by \cite{Par:63}
to describe the solar wind. Although a complete treatment of the evaporative
wind 
requires the integration of the energy, mass, and momentum transfer equations,
we can estimate the outflowing particle flux $\Phi_{H}$  (e.g. \citealt{Wat:81}) by
equating the energy input $(\epsilon L_{\rm xuv}/4) \times (R_1/r)^2 $ to
the energy required for hydrogen to escape $GM_p m_H/R_p$, giving 
\begin{equation}
\Phi_{H}\simeq\frac{\epsilon L_{\rm xuv} R_1^2 R_p}{4 r^2 G M_p m_H}
\label{loss}
\end{equation}
\noindent
where $R_1$ is the planet's radius where most of the XUV radiation is absorbed,
defined as the level where the optical depth is unity, $r$ is the orbital
distance, $R_p$ and $M_p$ are the planet's radius and mass respectively and
$ \epsilon L_{\rm xuv}$ the fraction of the stellar XUV luminosity (see
$\S3.2$) that is converted into thermal energy.  A proper
determination of $\epsilon$ requires the solution of the atmospheric structure
\citep{Yell:04}. Most calculations in the literature assumed the value of
$\epsilon = 0.63$ determined for Jupiter by \cite{Waite:83} (e.g.
\cite{Yell:04}). Note, however, that \cite{Cp:06} have recently
estimated the X-ray contribution to exospheric heating in hydrogen-rich
planetary atmospheres and found heating rates significantly higher than
those used by \cite{Yell:04} for planets at the same distance from the
central star. Given the high effective temperatures reached by PN central
stars we consider it reasonable to assume a value of $\epsilon$ = 1.

The determination of $R_1$, or ~$\xi = R_1/R_p$, is not an easy task, 
as it requires the full solution of the hydrodynamical escape problem with
radiation transfer in a strongly externally heated atmosphere. \cite{Wat:81}
assumed a single layer atmosphere and an iterative method to solve the
problem. \cite{Letal:03} and \cite{Betal:04} used Watson's solution for $R_1$
which matched the observed expanded atmosphere of HD 209458b ($R_1$ = 3
$R_J$). Note that in this approach it is implicitly assumed that $R_1$ 
lies below the {\it sonic level}.  

Lower mass-loss rates (by a factor of 20)  than those estimated under
hydrodynamical blow-off for HD209458b, have been obtained by
\cite{Letal:04}, by using what they refer to as {\it geometrical blow-off}:
a combination of Jeans escape rates and tidal forces.
Under geometrical blow-off the high exospheric temperatures push the 
position of the exosphere to the Roche-lobe limit 
before hydrodynamical blow-off  (or energy-limited mass-loss) conditions
develop, resulting in lower mass-loss rates than those estimated under
hydrodynamical blow-off.  \cite{Tetal:05} found an
escape rate 16 times smaller (for a heating  efficiency of 0.6) than the
maximum evaporation rate derived by \cite{Letal:03} and \cite{Betal:04}  for
HD 209458b and consistent with that determined by \cite{Letal:04}. More
recently,  \cite{Jar:05} have shown that exoplanets 
at small orbital distances will be subjected to either hydrodynamical or
geometrical blow-off conditions depending on where the position of the
exobase reaches the Roche lobe limit. 

In the \cite{Tetal:05} simulation the
hydrodynamical escape of hydrogen from a planetary atmosphere was studied by
allowing 
transonic solutions and using a two dimensional energy deposition layer.
With this approach they found the escape rates to depend on the position of
the heating layer as 
well as on the amount of energy input, with the escape rate exponentially
increasing with the heating rate. \cite{Tetal:05} also
found Watson's escape fluxes to be too high when low density
hydrogen atmospheres are involved (as in the case HD 209458b). At
hydrogen number densities higher than 10$^{14}$ cm$^{-3}$,
\cite{Tetal:05} find higher scape rates than Watson.
Generally, the simulations produce higher escape rates than the ones 
given by Eq.~\ref{loss} for $\xi$ = ~3 as the density increases. This is the
result of the higher total amount of energy  absorbed in an extended
atmosphere, as opposed to that absorbed in a single layer (the approximation
used to obtain Eq.~\ref{loss}).

The mass-loss rates (obtained by using $\xi~=$~3 in  Eq.~\ref{loss})
from a 1~\Mj~planet for central stars (at  36\,000 \,K) with different masses
are plotted in Fig.~\ref{hydro}, versus the orbital distance. Note that the
value of $\xi~$ 
used in Fig.~\ref{hydro} is always smaller than that obtained 
by \cite{Erk:06} for the exoplanets in which the mass, radius and stellar
parameters are known. 
The different
lines account for the different central star masses considered, with the
mass-loss 
rates increasing with the stellar mass. 

It is important to note that for the {\it geometrical blow-off}
\citep{Letal:04} to operate, the planet has to be in a very close orbit to the
parent star which is never the case for planets to be found around PN central
stars.  Moreover, none of the calculations in the literature includes heating
rates as 
high as the ones expected for a planet orbiting a PN central star, and as it
has been shown by \cite{Cp:06}, the inclusion of X-ray irradiation from the
star strongly increases the heating in planetary exospheres.
Therefore, the mass-loss rates given in Fig.~\ref{hydro} should be considered
merely as order of magnitude approximations to the
actual mass-loss rates from a planet exposed to a PN central star. In addition,
it is very likely that the planet will inflate as radiation is transformed into
heat inside its atmosphere, which will lead to further increase in the
planet's evaporation rate with our approach.  An appropriate determination 
of the escape rate will require a solution of the hydrodynamic escape
equations for the case under consideration.

\section{THE PLANET EVOLUTION UNDER MASS-LOSS}

Central stars of PN do not maintain high  luminosities for an
extended period of time and therefore the evaporation rates shown in
Fig.~\ref{hydro} will change as the star evolves. In Fig.~\ref{hydro_time} we
show the evolution of the evaporation rate of a 1\Mj~planet orbiting a 0.56
\Mso~PN central star at two orbital distances: 1.5\,AU and 5\,AU (solid and
dashed lines respectively). The evaporation rate has been computed using
$\xi=3$ in Eq.~\ref{loss} and the evolution of the stellar flux from the models
of \cite{Vw:94} extended to the white dwarf cooling tracks using the
\cite{Pm:02} models. 

It is important to note that the evolution of the evaporation rate shown in
Fig.~\ref{hydro_time} is clearly an oversimplication of the real case, which
involves factors other than the evolution of the stellar luminosity. For
instance, feedback processes are important because the evaporation  rate
depends on the planet's  structure which is itself determined by the planet's
reaction to evaporation and heating.  In order to establish the importance of
the feedback processes in determining the evaporation rates we need an
estimate of the 
relevant timescales. The timescale over which significant reaction takes place 
is given by the thermal timescale $\tau_{\rm th}$. For
the case under consideration the heating by radiation is taking place in a
thin  layer near the surface of the planet where the opacity is very high.
Therefore, the relevant thermal  timescale is the one
associated to the heating of the planet's external layer by an external heating
source, $\tau_{\rm th}=GM_{\rm p} \Delta M/2RL_*$, where $\Delta M$ is the 
mass that is being heated and  $L_*$ is the stellar luminosity
arriving at the planet's surface (multiplied by a correction factor that
accounts for the heating efficiency). Since Jupiter-like planets are believed to
be mostly convective, with a  radiative  external zone that contains $\approx$
0.03\% of the mass \citep{Gui:96},  it is reasonable to assume that the stellar
energy is mostly deposited in this layer, which then gives 
$\tau_{\rm th}\approx 10^6$ \,yr 
(for a 0.56 \Mso~star). 
We should now compare  $\tau_{\rm th}$ with the mass-loss timescale
$\tau_{\dot{M}}$, $\tau_{\dot{M}}~=~M_{\rm p}/\dot{M_P}$, and if we
assume that $\dot{M_p}$ = $\Phi_{\rm H}$ as given by Eq.~\ref{loss} then
$\tau_{\dot{M}}\approx 10^5$\,yr (for 1 \Mj~and
0.56 \Mso). Therefore, for  range of orbital distances, 
$\tau_{\dot{M}} \le \tau_{\rm th}$ which implies that the planet's mass is
decreasing faster 
than the time the planet has to readjust to a new thermal equilibrium. 

The
planet has two ways to adjust to a new thermal equilibrium: to increase its
radius or to increase its effective temperature. If the planet reacts to the
external heating by increasing its effective 
temperature, then the evolution of the mass-loss rates computed using
Eq.~\ref{loss} and the time-dependent stellar flux are a good first order
approximation to the actual evaporation rates. However, given the high
mass-loss rates involved, the planet may react by increasing its radius on a
timescale given by 
$\tau_{\rm th}$, in which case the mass-loss rates will be larger than
estimated.  In fact, 
once the mass-loss 
timescale becomes much shorter than the readjustment (thermal) timescale
($\tau_{\dot{M}} \lesssim 0.1 \tau_{\rm th}$), a
runaway mass-loss may ensue (e.g. \citealt{Betal:03}).

\subsection{Other Evaporative Processes}

There are other non-thermal processes that might contribute to the planet
evaporation, e.g. photodisociation and ram pressure stripping. An increase in
the planet's radius will  decrease the density of the outer layer allowing the
ionization front to propagate inwards, deeper into the structure of the planet.
It can be easily shown that given the high densities involved,
photoioniziation in
itself cannot destroy a Jupiter-like planet. However, the ionization of the
gas increases  the  pressure (as it increase the number of particles and the
temperature), causing an expansion of the outer layers that is driven from the
outside and not from the luminosity of the planet's interior. It has even
been suggested that a Jupiter like planet could be detected by the 
variable hydrogen recombination line emission it would emit in its outer
atmosphere which is photoionized by the PN central star \citep{Cetal:01}.

Ram pressure stripping caused by the stellar wind might become a very
efficient process in removing mass from the outer layers, once the planet
expands. Ram pressure stripping could
dominate during the early stages of the evolution of the PN phase, when high
velocity winds (with still relatively high densities) directly impact the
planet's surface (at that stage $\rho_{w} v_{w}^2~|_{star} >
(\dot{M_p}/4 \pi R_p^2) v_e~|_{planet}$ with $v_e$ the planet's scape
velocity). Ram   
pressure stripping due to the stellar proper motion has been shown to be a
very efficient process in removing mass from PN shells \citep{Vgm:03,Vs:05}.
Furthermore, evaporation rates might increase significantly when one takes
into account the rotation of the planet \citep{Bur:69}. 

As the star evolves, the wind velocity increases, reaching a velocity above
which the shocked gas is not able to cool down radiatively. An adiabatic shock
then develops at the interaction region between the high-velocity stellar wind
and the dense material ejected previously during the AGB, forming what is known
as the hot bubble. The post-shocked gas temperature can easily reach 10$^6$--
10$^7$\,K and has been observationaly confirmed through the detection of
diffuse extended X-ray emission \citep{Chu:01,Ggc:02,Getal:05}. The outer
radius of the hot bubble can reach $\sim$0.1 {\rm pc} and the reverse shock
region is expected to be very close to the star \citep{Vmg:02}. Therefore, the
planet's orbit will be immersed inside an environment at the hot bubble
temperature of 10$^6$-10$^7$\,K. The hot bubble, however, does not develop
inmediately \citep{Vmg:02}, because  the wind velocity needs to be high enough
to develop an adiabatic shock. It is then expected that the temperature of the
planet's atmosphere will be first set by the absorption of XUV radiation as
described in \S 3.2, and only during later stages of the star's evolution the 
hot bubble will provide an additional heating source.  

It can be expected that photoionization, ram pressure stripping and the
presence of the hot bubble will increase the
evaporation rate from a gas planet orbiting a PN central star. All these
processes have not been included in the evaporation rates given in
Fig.~\ref{hydro} (for $\xi=3$ which should therefore be regarded as
representing conservative lower limits.

\section{PLANET SURVIVAL}
We have shown that for small orbital distances the planet's structure will
change on timescales longer than the 
mass-loss timescales and that the most likely response will be an increase in
the planet's radius. As a first approximation and since we are
not computing the rate of change of the planet's radius (a problem that
requires a full hydrodynamic calculation) 
we have estimated the planet's survival by integrating the evaporation
rates obtained by evolving the stellar flux under different assumptions for
$\xi$.  As mentioned in \S4.1, $\xi$=3 represents the most conservative
scenario, most likely a lower limit to the evaporation rate of the
planet. Note that for this particular problem, an increase
in the planet's radius will naturally lead to lower  densities in the outer
layer, making the planet more vulnerable to the erosion caused by
photoionization and ram pressure stripping. Since we have not quantified all
these effects when estimating the evaporation rates in Eq.~\ref{loss},
using $\xi$=10 is probably a reasonable assumption.  Note that
\cite{Tetal:05} have shown that the position of the exobase for the extended
atmosphere of HD\,209458b is close to $\xi~=~10$.

Table~2 lists the percentage of the planet that is evaporated until a 0.56
\Mso~star enters the white dwarf cooling track. Column (1) gives the  initial
planet mass, column (2) the orbital radius, and columns (3), (4) and (5) give
the percentage of the  planet mass evaporated by using $\xi$=3, 5, and 10
respectively. Note that although we have extended the evolutionary time into
the white dwarf cooling sequence, most of the planet's evaporation takes place
as the star is evolving during the PN phase.  

In Fig.~\ref{final_fig} we show the region (on the planet mass versus orbital
distance plane) inside which Jupiter-like planets will be destroyed, 
as the star evolves off the main sequence. The left and right panels
are for planets orbiting a 0.56 and 0.9 \Mso~star respectively (which
correspond to 1 and 5 \Mso~main sequence masses). The light gray area
represents the region   inside which the planet will be engulfed and most
likely destroyed as the star ascends the AGB. The dark gray and
shaded areas (computed for $\xi$=3 and 10 respectively) represent the regions
for which planets will lose 50 \% of their mass before the star enters the
white dwarf cooling track due to the evaporation caused by thermal heating. 
Note that the region for which the planet will undergo total evaporation due
to thermal heating is inside the AGB stellar radius for the 0.9 \Mso~star.

Giant planets are thought to be formed in cool regions of the protoplanetary
disk--beyond the ice line-- where there is enough solid material to produce a
core that will capture the gas \citep{Sl:00,Ketal:04}. Planet
migration (e.g. \citealt{Tetal:98,Arm:02}) is then used in order to explain the
close distances at which such planets are observed around sun-like
stars (e.g. \citealt{Mq:95,Vm:04,Metal:05}).  Most of the planets recently
discovered are giant planets orbiting sun-like stars at orbits $\le$ 5 \,AU.

The conditions for planet survival as the star evolves off the main sequence
depend on the initial mass of the star. From Fig.~7 we see that most of the
close-in planets (r $\le$ 1.5 \,AU) will be destroyed as they get engulfed by
the star during the 
AGB phase (see Table~1 for a list of the maximum stellar radius reached during
the AGB versus the stellar mass). As the star leaves the AGB and enters the PN
phase, high effective temperatures at very high luminosities set up an
evaporation flux at the  surface of the planet. At certain orbital distances
the evaporation rates are high enough to cause a total destruction of the
planet. By integrating the evaporation rates as the star evolves during the PN
into the white dwarf cooling track, we find that Jupiter-like planets will be
destroyed if they remain at orbital distances \,r$\le$ 5\,AU from a low mass
white dwarf (M$_{\rm WD} \le$ 0.63 \Mso) and large planet ablation is expected
up to 10\,AU. In particular, Jupiter in our own Solar system is barely
expected to survive. More massive stars evolve very fast during the  PN phase and do
not maintain high evaporation rates long enough to cause planet destruction,
unless the planets were to be found at small orbital distances (r$\le$2.5
\,AU). 
However, we have shown that planets orbiting the more massive PN central stars
cannot be found at small orbital distances: if a planet orbiting a 0.9 \Mso
progenitor is to survive AGB   engulfment then its orbit has to be at r$\ge$
29 \,AU. 

Over the last few years several groups have surveyed white dwarfs in the
search for planets with no positive results reported so far. 
The wide field proper motion survey around 261 white dwarfs by \cite{Farihi:05}
was sensitive to 100 to 5000 \,AU with masses down to 52 \Mj, and their deeper
near field search of 86 white dwarfs could detect 10 \Mj~at separations between
50 and 1100 \,AU. \cite{Mull} have recently surveyed 124 white dwarfs
with Spitzer and found no evidence for planets with M $\ge$ 5
\Mj. Using direct imaging \cite{Zinn} have ruled out the presence of giant
planets (6-12 \Mj) around the 7 known white dwarfs of the Hyades cluster. 
A planet (4 \Mj) was discovered  in the binary system Gliese 86 by
\cite{Mandm} orbiting the K1 dwarf in Gliese 86A, which has a 0.55
\Mso mass white dwarf companion, Gliese 86B, at 21 \,AU. 
At this distance from
the white dwarf, we expect the planet to have been losing mass during the
PN phase at such a low rate that it would not have affected its survival.

On the other hand, three white dwarfs are known to posses an infrared excess, 
GD~362 \citep{Zb:87,Beck:05,Kilic:05}, G29-39 \citep{Jura:03} and the
recently discovered GD\,56 \citep{Kilic:06}. To explain G\,29-38
\cite{Jura:03} proposed a disk formed by the tidal destruction of an
asteroid. The destruction and accretion of a planetary body continues
to be the widely accepted scenario to explain the high anomalous photospheric
abundances in the massive white dwarf GD\,362, where a massive
accretion disk has been detected \citep{Beck:05,Kilic:05}. 
We have shown that 
planets around the most massive white dwarfs are expected to be found at
r$\ge$ 30 \,AU, due to the expansion of the planetary orbit during AGB
mass-loss. How a  
planet at such a large distance from the
star would then be disrupted (in order to form an accretion disk) is by no means
clear. Another more plausible scenario for the presence of accretion disks
around massive white dwarfs, is that they are formed by the merger of two white
dwarfs \citep{Liv:05}, a scenario that can also explain the presence of
planets around massive white dwarfs at small orbital distances.

\section{CONCLUSIONS}
Planets at orbital distances within the reach of the
stellar radius during the AGB phase will spiral-in  and totally
evaporate. The AGB stellar radius depends on the stellar mass and therefore so 
does the distance from the parent star at which a planet is destroyed, r$\le$
1.6 \,AU for $M_*$=0.56 \Mso and r $\le$5.3 \,AU for $M_*$=0.9 \Mso.

If the
planet avoids engulfment its orbit will typically expand. The final
masses of white dwarfs are primarily established during 
the AGB  phase due to heavy mass-loss. The more massive progenitors (when on the main
sequence) lose larger amounts of mass during the AGB, thus the planets orbiting
these stars will experience the largest orbital readjustments. 

We show that as the star evolves into the PN phase, high temperatures are
reached in the outer atmosphere of the planet resulting 
in the development of blow-off conditions.  As the star evolves through 
the PN phase into the white dwarf cooling track:

\begin{enumerate}
\item An general outflow is established at the surface of a
  substellar object. The evaporation flux decreases with the orbital distance,
  and as the star evolves, with higher evaporation fluxes expected
  for planets orbiting the more massive stars. The evaporation rates are
  sufficiently high  (10$^{-5}$ \Mj yr$^{-1}$ at 5\, AU for a 1\Mj~planet)
  that these evaporating planets may be detected using spectroscopic
  observations, similar to those performed for HD 209458b by  \cite{Vm:03}.

\item Under these evaporation rates, planets with $M_p\le$1 \Mj~will not
  survive the PN phase if located at orbital distances 
  $r\le$ (3--5)\,AU.  

\item Planets with  $M_p>$ 2\Mj~survive the PN phase down to orbital distances
  of $\sim$ 3\,AU around low-mass  ($M_{\rm WD} \sim$ 0.56 \Mso) central
  stars.

\item Planets around white dwarfs with masses $M_{\rm WD} \gtrsim$ 0.7
  \Mso~(that formed from single stars with masses $M_{MS} \gtrsim$ 2.5
  \Mso~), are generally expected to be found at orbital radii r $\gtrsim$ 15
  \,AU due to the effects of mass-loss on the AGB phase. If planets are found
  at smaller orbital radii around such massive 
  white dwarfs, they had to form as a result of the merger of two white
  dwarfs \citep{Liv:05}.

\end{enumerate}

\noindent
{\bf Acknowledgment:} We are grateful  to James Herald for providing us with
the stellar fluxes from the 
CMFGEN code.

\begin{figure}
\plotone{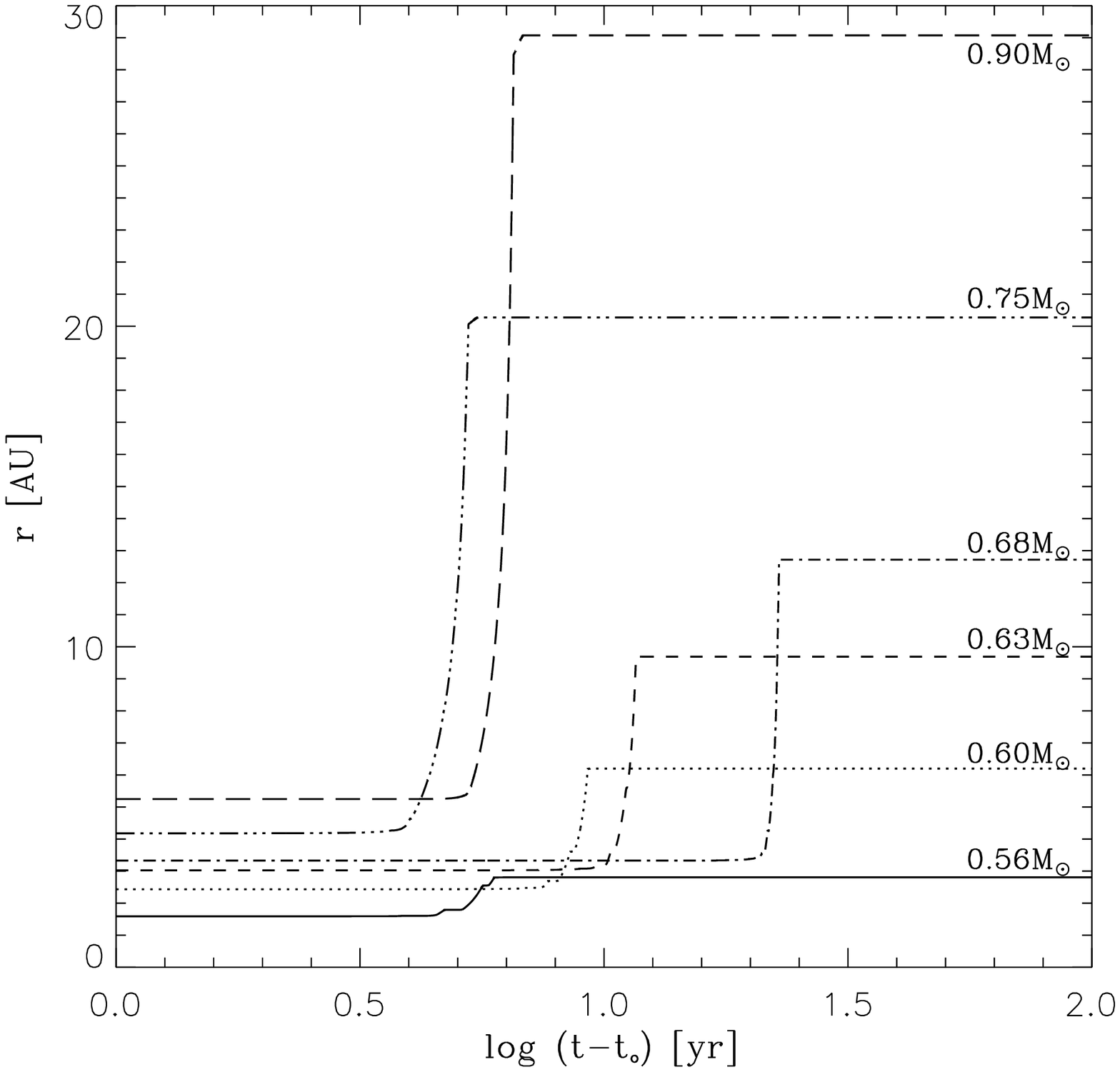}
\caption[ ]{
Evolution of the orbital distance caused by the stellar mass-loss during the
AGB phase. The time is given 
in logarithmic scale, with t$_o$ representing the begining of the AGB
evolution. Each curve represent a different PN central star
mass, marked on each curve and with different line-styles. We have used an
initial 
orbital distance that equals the maximum radius reached by the star during the
AGB phase.
\label{f1.eps}}
\end{figure}

\begin{figure}
\plotone{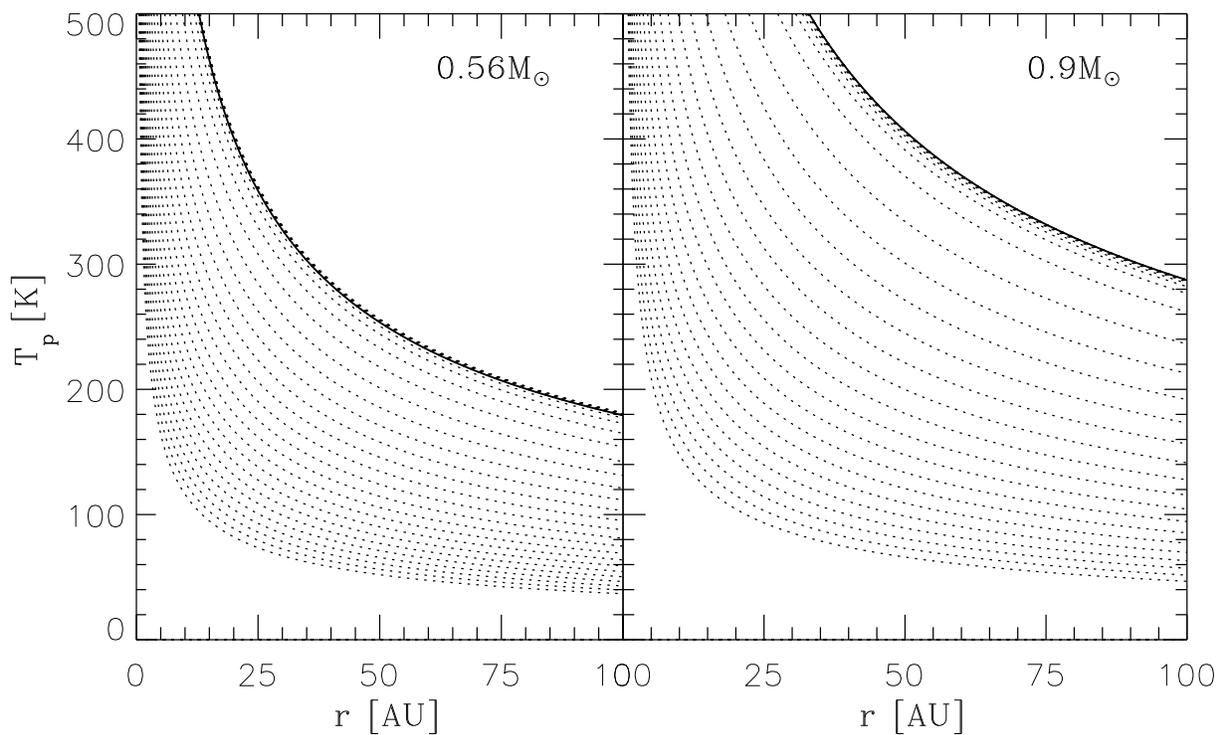}
\caption[ ]{
Left: the solid line represents the planet's equilibrium temperature (from
Eq.~\ref{Teq}) versus orbital distance for a star of 0.56 \Mso~as it enters
the PN phase. The dashed lines follow the evolution of the star 
up to 1.1$\times10^6$ \,yr after the PN phase starts. The lowest
equilibrium temperature corresponding to the latest point in the evolution.
Right: the same for a 0.9 \Mso~star evolved to 1 $\times10^5$ \,yr into the
PN phase.
\label{teq_fig}}
\end{figure}

\begin{figure}
\plotone{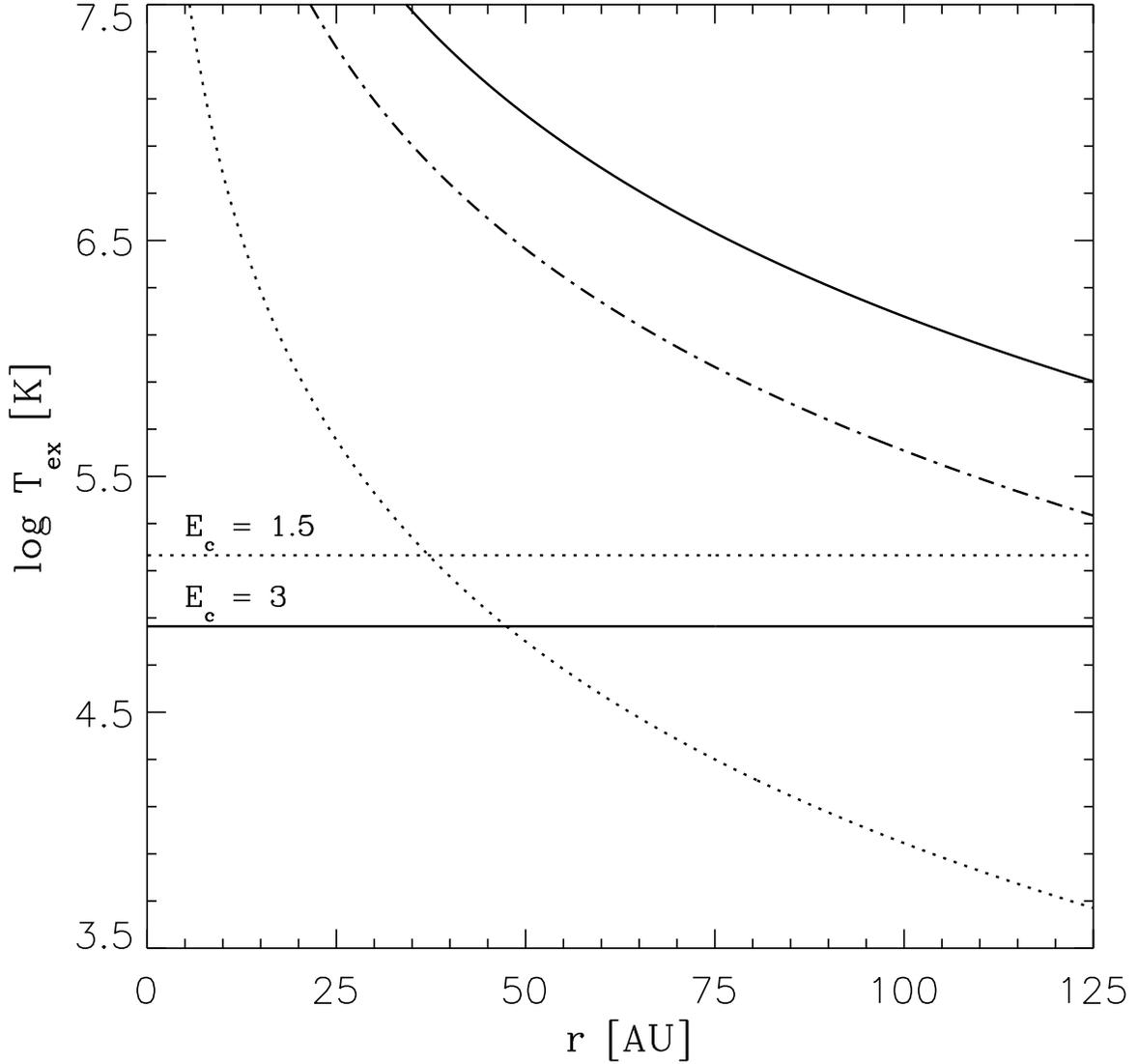}
\caption[ ]{
Exospheric temperature (in Logarithmic scale) reached at the atmosphere of a
1 \Mj~planet orbiting a 0.56 \Mso~star (1 \Mso~main sequence) plotted
versus the 
orbital distance. The different curves (solid, dashed, and dotted) are the
result of changing the XUV intensity due to the evolution of the star and are
chosen at luminosities of 3.6, 1.4 and 0.1 $\times 10^3$ \Lso~respectively
which correspond to stellar effective temperatures of 36\,000, 115\,000 and
96\,000 \,K respectively. The  solid horizontal line represents
$T_{ex}$, or the escape parameter $E_c$=3 (as defined in the text), above which the
atmosphere reaches blow off conditions.  The position of  $E_c$=1.5 is also
shown as a dotted line.
\label{exo_temp}}
\end{figure}

\begin{figure}
\plotone{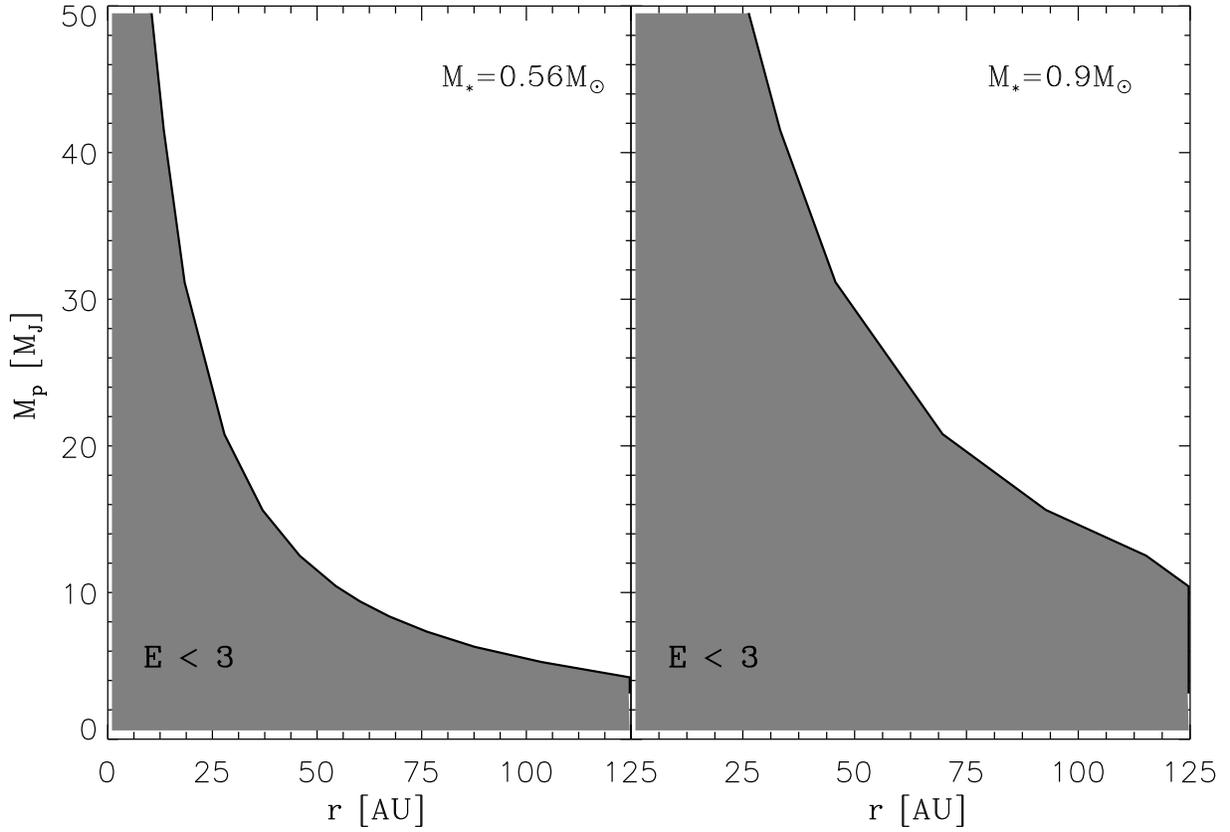}
\caption[ ]{The grey area represents the parameter space region (planet's
  mass versus orbital distance) for which the planet's atmosphere reaches
  blow off conditions using a critical escape parameter, $E_c$~=~3. Left: $E_c$
  computed for  planets embeded in the PN environment
  created by the lowest mass PN progenitor 0.56 \Mso~(1 \Mso~main sequence mass)
  at an early age, when $L_*$=3.6$\times 10^3$ \Lso~and $T_{eff}$ = 36\,000
  \,K. Right: the same but for a 0.9 \Mso~central star.
\label{exo_mass}}
\end{figure}

\begin{figure}
\plotone{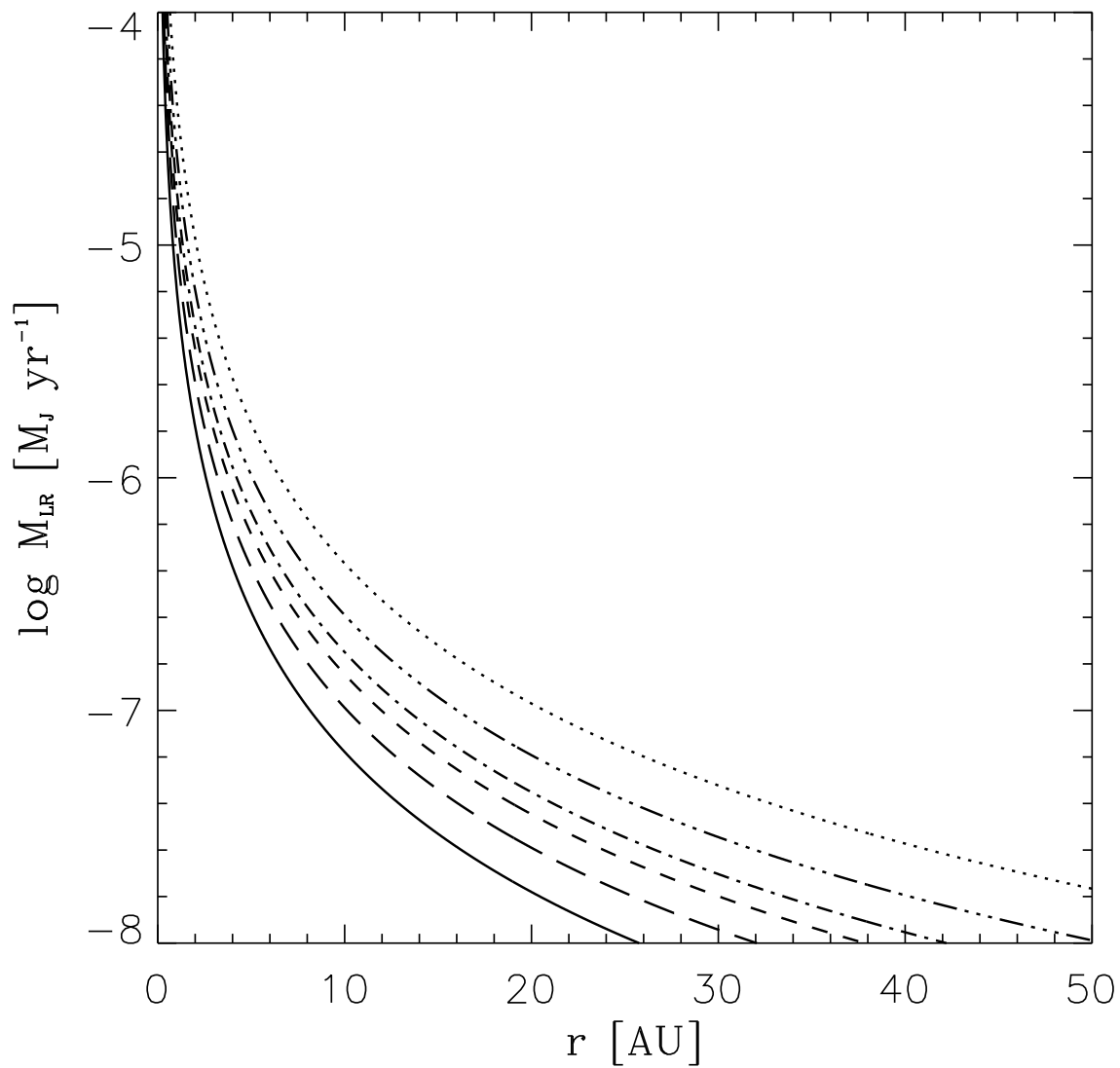}
\caption[ ]{
Mass-loss using $\xi=3$ (see text) (in logarithmic scale and M$_{\rm J}$
yr$^{-1}$) versus orbital distance of 
a Jupiter-like planet 
under hydrodynamic escape conditions. The solid line is for the 
radiation field emited by the lowest mass core 30\,000 yr after the star
enters  the PN phase of evolution. The different lines are for all the other
stellar masses 
considered with the radiation (and therefore the mass-loss rates) increasing
with the stellar mass. 
\label{hydro}}
\end{figure}

\begin{figure}
\plotone{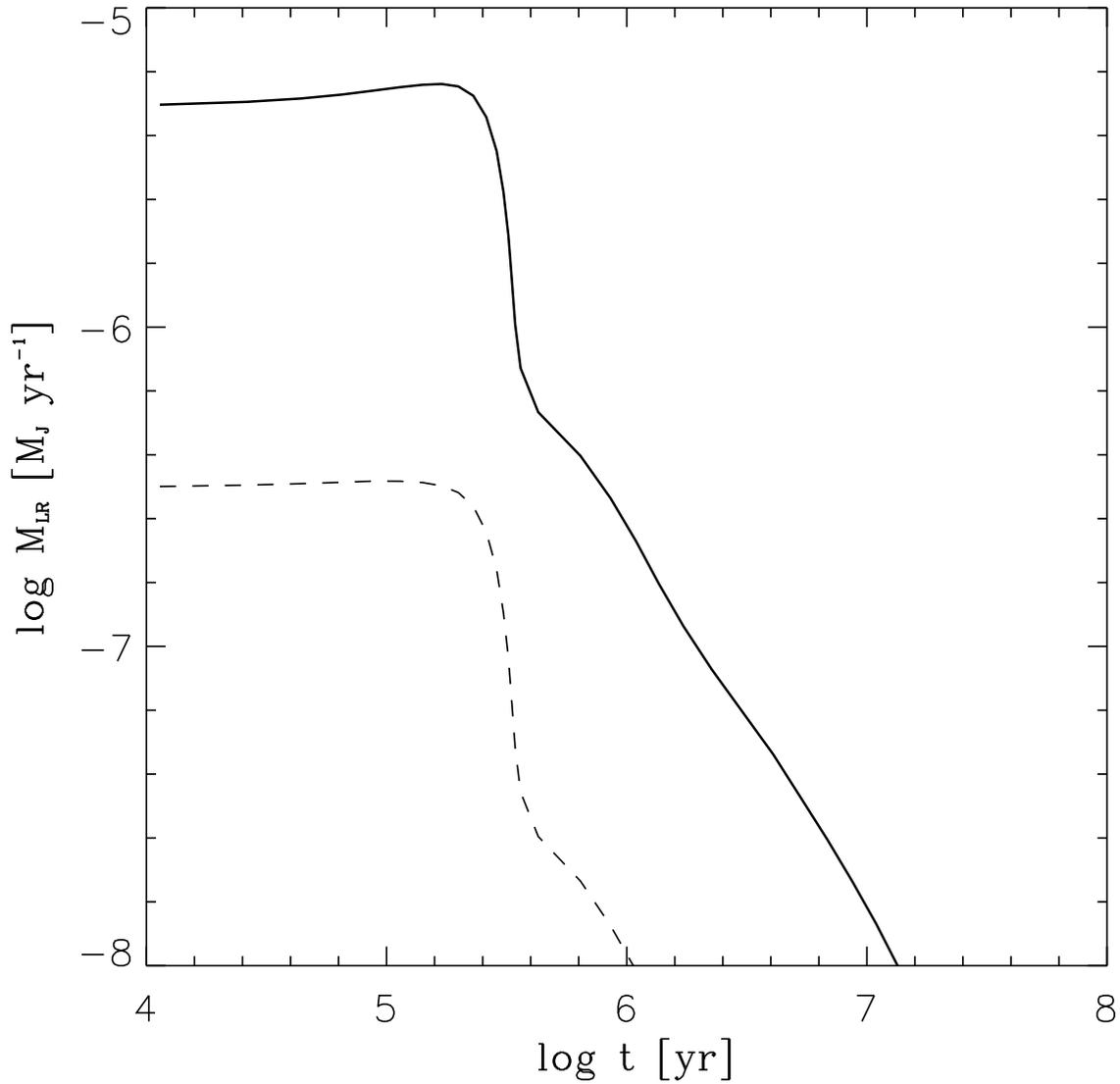}
\caption[ ]{
The solid line represents the evolution of the evaporation rate (in logarithmic scale and M$_J$
yr$^{-1}$) of a 1 \Mj~mass planet 
under hydrodynamic escape conditions and at a distance of 1.5 \,AU  from a .
 for a 1
\Mso star. The same for the dashed line but for a distance of 5 \,AU. Time zero is
set as the star leaves the AGB phase .  The
evolution of the stellar luminosity during the PN phase has been taken from
the stellar evolutionary models of \cite{Vw:94} and it has been extended to
the white dwarf cooling sequence using the models of \cite{Pm:02}.
\label{hydro_time}}
\end{figure}

\begin{figure}
\plotone{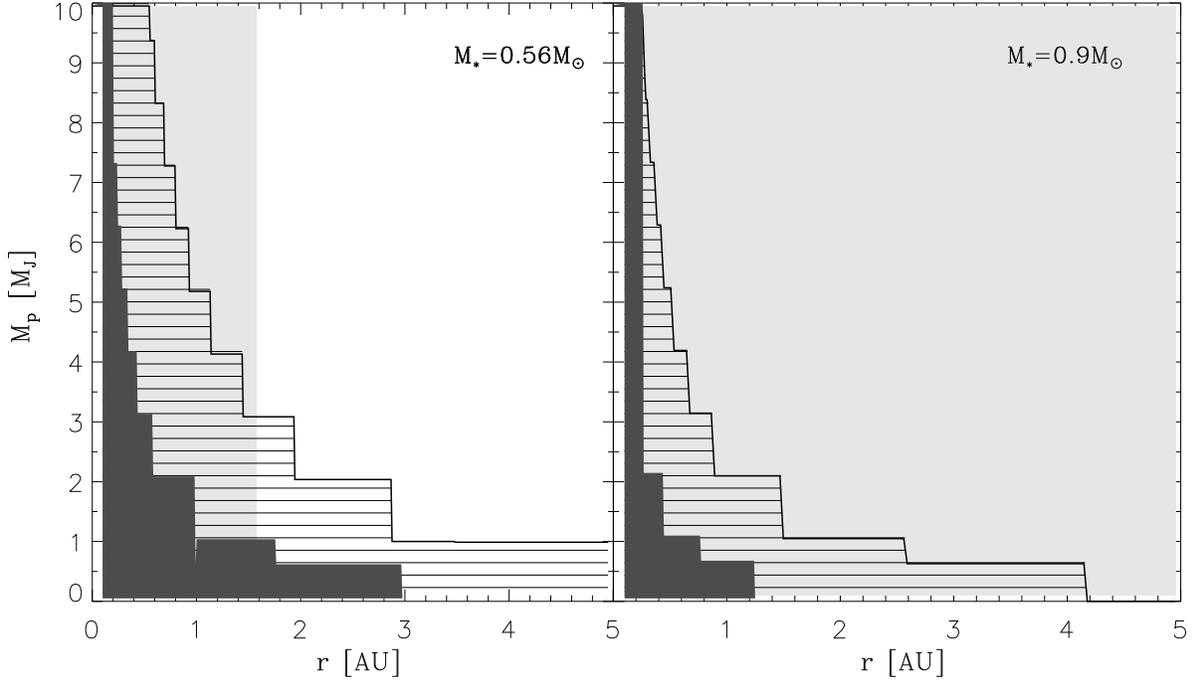}
\caption[ ]{Left: The dark gray area represent the region where
  a planet loses 50 \% of its mass when orbiting a 0.56 \Mso~star by using
  $\xi=3$ (see text). The same for the region with parallel line filling but
  using 
  $\xi=10$.  The light gray area
  represents the orbital distance reached by the stellar radius during the
  AGB phase. Right: the same for planets orbiting 0.9 \Mso~central star.
 \label{final_fig}}
\end{figure}

\begin{deluxetable}{cccc}
\tabletypesize{}
\tablecaption{MASSES, ORBITAL DISTANCES}
\tablewidth{0pt}
\tablehead{
\colhead{M$_{MS}$} & \colhead{M$_{WD}$} & \colhead{R$_*^{max}$} &
  \colhead{$r(t)/r_o$}\\
\colhead{[M$_{\rm \odot}$]} & \colhead{[M$_{\rm \odot}$]} & 
\colhead{[\,AU]} & \colhead{ }
}
\startdata
1.0  &  0.56 & 1.59 & 1.8 \\
1.5  &  0.60 & 2.43 & 2.5 \\
2.0  &  0.63 & 3.02 & 3.2 \\
2.5  &  0.68 & 3.33 & 3.8 \\
3.5  &  0.75 & 4.18 & 4.9 \\
5.0  &  0.90 & 5.25 & 5.5 \\
\enddata
\end{deluxetable}

\begin{deluxetable}{ccccr}
\tabletypesize{}
\tablecaption{\% EVAPORATION FOR 0.56 \Mso~STAR}
\tablewidth{0pt}
\tablehead{
\colhead{M$_p$[\Mj]} &\colhead{r[AU]} & \colhead{$\xi$=3} &
\colhead{$\xi$=5} & \colhead{$\xi$=10}}
\startdata
1    &1 &  100\,\% & 100\,\% & 100\,\% \\
     &2  &  30\,\%   & 100\,\%  & 100\,\%  \\
     &3  &  20\,\%  & 40\,\%  & 100\,\%  \\
     &4  &  10\,\%  & 20\,\%  & 100\,\%  \\
     &5  &  5\,\%  & 20\,\% & 70\,\% \\
     &10 &  1\,\% & 5\,\% & 20\,\% \\
\hline
5    &1 &  4\,\% & 10\,\% & 52\,\% \\
     &2  & 1\,\% & 3\,\%  & 10\,\%  \\
     &3  & 0.4\,\% & 1\,\% & 5\,\%  \\
     &4  & 0.2\,\% & 0.6\,\%  & 3\,\%  \\
     &5  & 0.1\,\%  & 0.4\,\% & 1.6\,\% \\
     &10 & 0.04\,\% & 0.1\,\% & 0.4\,\% \\
\enddata
\end{deluxetable}

\end{document}